\documentclass[showpacs,preprintnumbers,amsmath,amssymb]{revtex4-1}
\usepackage[dvips]{graphics}
\usepackage{color}

\begin{document}
% \draft
\title{Photon-number entangled states generated in Kerr media with optical parametric pumping}
\author {A. Kowalewska-Kud{\l}aszyk}
\email{annakow@amu.edu.pl}
\affiliation{Nonlinear Optics Division, Department of Physics, Adam Mickiewicz
 University, Umultowska 85, 61-614 Pozna\'n, Poland}
%\date{\today}
\author {W. Leo\'nski}
\email{wleonski@proton.if.uz.zgora.pl}
\affiliation{Quantum Optics and Engineering Division, Institute of Physics, University of Zielona G\'ora, Prof.~Z.~Szafrana 4a, 65-516 Zielona G\'ora, Poland}
\author{Jan Pe\v{r}ina Jr.}
\affiliation{Palack\'{y} University, RCPTM, Joint Laboratory of Optics,
  17. listopadu 12, 771 46 Olomouc, Czech Republic} 
%Joint Laboratory of Optics of Palack\'{y} University and Institute of Physics of Academy of Sciences of the Czech Republic,
%17. listopadu 50a, 772 07 Olomouc, Czech Republic}
\email{perinaj@prfnw.upol.cz}

\date{\today}

\begin{abstract}
Two nonlinear Kerr oscillators mutually coupled by parametric
pumping are studied as a source of states entangled in photon
numbers. Temporal evolution of entanglement quantified by
negativity shows the effects of sudden death and birth of
entanglement. Entanglement is preserved even in asymptotic states
under certain conditions. The role of reservoirs at finite
temperature in entanglement evolution is elucidated. Relation
between generation of entangled states and violation of
Cauchy-Schwartz inequality for oscillator intensities is found.
\end{abstract}
\maketitle

\section{Introduction}

Generation of entangled states and their protection from
disentanglement is a crucial task in quantum information
processing theory. Entangled states can be generated in various
physical systems. Entanglement can be observed in various degrees
of freedom. The most common systems considered as sources of
entangled states are spin systems (NMR), quantum dots\cite{LV98},
trapped ions \cite{CZ95}, double-well Bose-Einstein condensates
\cite{VPA04}. Typical constituents of these systems can be
represented by two (or more) level atoms that mutually interact
and create this way entangled states. We speak about systems
composed of qubits (see for example \cite{F10} and the references
quoted therein), qutrits or higher dimensional qudits. On the
other hand, entangled states can be found also in field systems,
namely in optical fields. Entanglement emerges here due to mutual
interaction among modes of optical fields. This interaction
typically occurs in nonlinear optical processes \cite{Boyd2003}.
Entanglement can then be observed in polarization states (for
example \cite{PHJ94}) or photon-number states
\cite{Bouwmeester2000}.
Namely states entangled in photon numbers are promising for the near future due to the progress
in construction of photon-number resolving detectors in the last years. Their construction can be based on
optical-fiber loop interferometers \cite{Rehacek2003,Achilles2004,Fitch2003}, absorption
in superconductor wires \cite{Miller2003}, use of intensified CCD cameras \cite{Haderka2005,Hamar2010},
complex semiconductor structures \cite{Kim1999} or hybrid photodetectors \cite{Bondani2009}.

Every real physical system interacts with its environment
\cite{Fick1990}. Such interaction leads to losses of coherence,
and consequently, to gradual degradation of entanglement. In
quantum information processing this represents one of the crucial
problems. That is why many methods for coping with this problem
have been developed; quantum error correction
\cite{S95,EM96,G96,CRSS97}, decoherence free subspaces
\cite{PSE96,DG98,LW03}, proper preparation of qubit states
\cite{ZR98}, multiple system-decomposition method \cite{JD08}, to
name few. It is well known that an entangled state interacting
with a zero temperature Markovian environment asymptotically
decays in time. However the character of zero-temperature
reservoir decay can be altered. Under specific conditions
disentanglement may suddenly occur (sudden death of entanglement)
\cite{ZHHH01,D03,YE04}. Moreover, if a system interacts with a
thermal or squeezed \cite{ILZ07} reservoir the entanglement is
lost in finite time as well. We should also mention that the
effect of sudden entanglement birth can also be observed
\cite{FT06,FT08,LRLSR08}.

In this paper, we consider two optical modes generated by
parametric down-conversion \cite{Mandel1995} pumped by an intense
optical field. The generated two optical modes propagate through a
medium characterized by Kerr nonlinearity that modifies their
statistical properties. States characterizing two optical modes
can be entangled in photon numbers (see for example \cite{PHJ94})
as a consequence of the fact that photons are emitted in pairs in
parametric down-conversion. Provided that interaction with
external reservoir is taken into account the effects of sudden
death and sudden birth of entanglement can be observed.
Entanglement will be quantified by \textit{negativity} function
and its presence will be compared with nonclassical behavior of
distributions of integrated intensities of two modes. We note that
experimental characterization of nonclassical two-mode optical
fields is experimentally available \cite{PKPBAA09}.

Optical fibers are suitable candidates for experimental generation
of the predicted entangled states. They allow both the generation
of photon pairs in parametric processes \cite{Li2005,Fulconis2005}
and exhibit Kerr nonlinearities \cite{Sizmann1999}. Nonlinear fibers allow to generate sufficiently high photon-pair
fluxes. For example, a nonlinear fiber $300$~m long embedded into
a Sagnac interferometer allows to generate up to $ 10^7 $ photon
pairs per $1$~mW of optical pumping \cite{Li2005}. Similar
photon-pair generation rates have also been reported from a 2~m
long micro-structured fiber \cite{Fulconis2005}.

However, we note that
nonlinear systems excited directly by an external coherent field
[referred as \textit{nonlinear quantum scissors}
\cite{ML06,KL06,KL09}] require giant Kerr nonlinearities that can
be achieved, e.g., using induced transparency
\cite{KZ03,BC08,SRKD08}.

The paper is organized as follows. A quantum model of two
interacting optical fields is formulated and solved in Sec.~II
that also contains considerations about entanglement.
Generalization of the model to the case of interaction with
reservoir is provided in Sec.~III. Nonclassical behavior of
integrated intensities of two modes is discussed in Sec.~IV.
Sec.~V provides conclusions.

\section{The model, its solution and entanglement evolution}

The considered system is composed of nonlinear medium
(characterized by Kerr nonlinearity) that is excited by an
external field of strength $g$. At quantum level, two photons
(belonging to different field modes) are emitted together in the
parametric process. Interaction Hamiltonian $ \hat{H}_{int} $ of
the system can be written in the following form:
\begin{eqnarray}
 \hat{H}_{\rm int} &=& \hat{H}_{\rm Kerr} + \hat{H}_{\rm par}
  \label{hamilt} \\
 \hat{H}_{\rm Kerr} &=& \frac{\chi_a}{2}(\hat{a}^\dagger)^2\hat{a}^2
  +\frac{\chi_b}{2}(\hat{b}^\dagger)^2\hat{b}^2,\\
 \hat{H}_{\rm par} &=&
  g\hat{a}^\dagger\hat{b}^\dagger+g^{\star}\hat{a}\hat{b}.
\end{eqnarray}
Here, Hamiltonian $ \hat{H}_{\rm Kerr} $ describes two nonlinear
Kerr oscillators (with nonlinearities $\chi_{a,b}$) whereas
Hamiltonian $ \hat{H}_{\rm par}$ means an effective interaction
Hamiltonian of the two-mode parametric process \cite{P91}. Model
Hamiltonians including both parametric and Kerr processes have
been analysed in \cite{PP00}. It should be noted that
our model does not include the Kerr cross term
$\hat{a}^\dagger\hat{a}\hat{b}^\dagger\hat{b}$, that is important
in some Kerr media (for discussion concerning models involving such Kerr cross term see, for example \cite{KP96,*KP97,CSKSL98} \textit{and the references quoted therein}). We note that the Hamiltonian analogous to that
considered here has also been used for example, by Bernstein
\cite{Bernstein93} or Chefles and Barnett \cite{CheflesBarnett96}.
The dynamics of the models excluding the Kerr cross term have also
been analyzed in other papers, for instance in
\cite{AbdelBaset00,Olsen06}.

It should be stressed out that various interesting effects beside
of those discussed in the quantum information theory can originate
in systems endowed with Kerr nonlinearities. For instance,
Miranowicz \textit{at al.} \cite{MTK90} have shown that discrete
superpositions of an arbitrary number of coherent states
(so-called Schr\"odinger cat-like or kitten states) can be
generated in these systems. Moreover, such nonlinear systems
(especially, nonlinear couplers with Kerr nonlinearities
\cite{ML06,KL06,KL09}) can serve as a source of maximally
entangled states (MES) useful not only in quantum optics but also
in other branches of physics. For instance, Kurpas \textit{et al.}
\cite{KDZ09} discussed nonlinear resonances and entanglement
generation in solid state models, particularly in those involving
SQUID systems.

At the beginning we neglect damping. That is why we simply apply
the Schr$\ddot{o}$dinger equation that can be written in the form
of a set of equations for complex probability amplitudes
$c_{nm}(t)$ corresponding to the Fock state with $ n $ photons in
mode $ a $ and $ m $ photons in mode $ b $. Assuming that the
initial state is the vacuum state in both modes
$|\Psi(t=0)\rangle=|0\rangle_a|0\rangle_b$ these equation can be
obtained in the form (we use units of $\hbar=1$):
\begin{eqnarray}   % 4
 i\frac{d}{dt}c_{nm}&=&\frac{1}{2}\left\lbrace n(n-1)\chi_a+m(m-1)\chi_b\right\rbrace c_{nm}
  \nonumber \\
 & & \mbox{} + g\sqrt{nm}\,\,c_{n-1,m-1} \nonumber \\
 & & \mbox{} + g^{\star}\sqrt{(n+1)(m+1)}\,\,c_{n+1,m+1}.
\label{ampl}
\end{eqnarray}
Inspection of the form of Eqs.~(\ref{ampl}) leads to the
conclusion that the dynamics of the system is not restricted to a
closed set of two-mode states. The number of states involved
considerably in the dynamics depends straightforwardly on the
level of excitation $g$ - the larger the value of $ g $ the larger
the number of involved states. Assuming the initial vacuum state
($|\Psi(t=0)\rangle=|0\rangle_a|0\rangle_b$) and lossless dynamics
only two-mode states having equal photon numbers in modes $a$ and
$b$ are involved in the dynamics.

To be more specific, the number of states that must be considered
depends on the value of ratio $r=g/(\chi_a+\chi_b)$ and length of
the time-evolution. For small values of $r<0.1$ the conditions
resemble those discussed in \cite{ML06,KL06,KL09} where
\textit{nonlinear quantum scissors} have been studied. In this
case, only states with small photon numbers in both modes $ a $
and $ b $ are involved. In particular, we can restrict the
dynamics of the system to just first three lowest states of the
whole set of two-mode states: $|0\rangle_a|0\rangle_b,
|1\rangle_a|1\rangle_b, |2\rangle_a|2\rangle_b$. Validity of this
approximation can be judged comparing the approximative solution
with the full (numerical) solution. Fidelity between the two
states has been found useful in this case. A typical example
investigated in Fig.~1 shows that the dynamics of the whole system
can be restricted to the subspace of three states with accuracy of
order $3\times 10^{-4}$.  For this case we assume that
$g=0.15$. This value seems to be the optimal one, since it is
small enough to keep the system's evolution closed within the
desired range of the states and to derive the analytical formulas
for the dynamics. Simultaneously, the value $g=0.15$ is sufficient
to observe the Bell-states generation. The reason is that if we
assume the value for $g$ to be too small, mostly the vacuum state
is populated and an efficient Bell-state generation is not
possible.

In this approximation, the evolution of probability amplitudes is
governed via the equations:
\begin{eqnarray}   % 5
 i\frac{d}{dt}c_{00}(t)&=&g^{\star}c_{11}(t)\nonumber\\
 i\frac{d}{dt}c_{11}(t)&=&gc_{00}(t)+2g^{\star}c_{22}(t)\nonumber\\
 i\frac{d}{dt}c_{22}(t)&=&2gc_{11}(t)+Ac_{22}(t),
\label{amp_smg}
\end{eqnarray}
% check
where $ A = \chi_a + \chi_b $.

The solution of Eqs.~(\ref{amp_smg}) can be obtained analytically:
\begin{eqnarray}  % 6
 c_{00}(t)&=&r_{01}\exp{(s_1 t)}+r_{02}\exp{(s_2 t)}+r_{03}\exp{(s_3 t)}\nonumber\\
 c_{11}(t)&=&r_{11}\exp{(s_1 t)}+r_{12}\exp{(s_2 t)}+r_{13}\exp{(s_3 t)}\nonumber\\
 c_{22}(t)&=&r_{21}\exp{(s_1 t)}+r_{22}\exp{(s_2
 t)}+r_{23}\exp{(s_3 t)}, \nonumber \\
 & &
\label{amp_smg_sol}
\end{eqnarray}
where
\begin{eqnarray}  % 7
 s_1 &=& \frac{ig}{3x}\left( -1+\frac{1+15x^2}{M}+M\right),
  \nonumber \\
 s_2 &=& \frac{g}{6x}\left(-2i-X_2/M\right)\, , \nonumber \\
 s_3 &=& \frac{g}{6x}\left(-2i+X_2^{\star}/M\right)\,.
\end{eqnarray}
Amplitudes $r_{ij}$ take the following form:
\begin{eqnarray} % 8
 r_{01} &=& \frac{12x}{m_1} \Bigl[-ix(7+75x^2)+K+M(-2ix+K) \nonumber \\
   & & \mbox{} +2ixM^2 \Bigr], \nonumber \\
 r_{02} &=& \frac{x}{m_2}\Bigl[ (-3-i\sqrt{3})(7+75x^2)x+(-3i+\sqrt{3})K \nonumber \\
   & & \hspace{-8mm} \mbox{} + 2(3-i\sqrt{3})xM+(3i+\sqrt{3})KM-4\sqrt{3}ixM^2\Bigr], \nonumber\\
 r_{03} &=& \frac{x}{m_3}\Bigl[(3-i\sqrt{3})(7+75x^2)x+(3i+\sqrt{3})K \nonumber \\
   & & \hspace{-8mm} \mbox{} -2(3+i\sqrt{3})xM+(-3i+\sqrt{3})KM-4\sqrt{3}ixM^2\Bigr], \nonumber\\
 r_{11} &=& \frac{12ixM}{m_1} \left(15x^2+(1+M)^2\right), \nonumber \\
 r_{12} &=& -r_{13}^{\star} = \frac{\sqrt{3}xM}{m_2}\left(X_2-4iM\right), \nonumber\\
 r_{21} &=& -\frac{72ix^2M^2}{m_1}, \nonumber \\
 r_{22} &=& \frac{12i\sqrt{3}x^2M^2}{m_2}, \nonumber \\
 r_{23} &=& r_{22}\frac{m_2}{m_{3}}.
\label{8}
\end{eqnarray}
Denominators occurring in Eqs.~(\ref{8}) for amplitudes are given
as:
\begin{eqnarray}  % 9
 m_1 &=& \left|X_1\right|^2, \nonumber \\
 m_2 &=& m_3^{\star} = \left(-1-15x^2+M^2\right)X_1.
\end{eqnarray}
Also the following parameters have been used in the above
relations:
\begin{eqnarray}  % 10
 X_1 &=& (3i-\sqrt{3})(1+15x^2)+(3i+\sqrt{3})M^2,\nonumber \\
 X_2 &=& (i-\sqrt{3})(1+15x^2)+(i+\sqrt{3})M^2,\nonumber \\
 M &=& \left(-1-9x^2+3ixK\right)^{1/3}, \nonumber \\
 K &=& \sqrt{3+66x^2+375x^4}, \nonumber \\
 x &=& g/(\chi_a+\chi_b).
\end{eqnarray}

The analytical solution in Eqs.~(\ref{amp_smg_sol}) valid for
three two-mode states allows to analyze entanglement in the
system. Here, we pay attention to \textit{qubit-qubit}
entanglement and quantify it using negativity.  This
entanglement measure is defined as~\cite{Peres,Horodecki96}:
\begin{equation}
N(\hat\rho)=\max\Big(0,-2\min_j \mu _{j}\Big), \label{negativity}
\end{equation}
where $\mu _{j}$ are the eigenvalues of the partial transpose of
the density matrix $\hat\rho^{\Gamma}$. The factor $2$ appearing
in this definition is chosen to get $N(\hat\rho)=1$ for Bell's
states. Negativity is a commonly used entanglement measure
allowing the entanglement quantification -- for maximally
entangled state we have $N(\hat\rho)=1$, whereas $N(\hat\rho)=0$
occurs for the cases when the entanglement disappears completely.
Moreover, it can be applied for more general models exceeding the
simple \textit{qubit-qubit} one.

In our system, we can define three subsystems of the qubit-qubit
form: the first subsystem is spanned by the states
$|0\rangle_a|0\rangle_b$ and $|1\rangle_a|1\rangle_b$ (negativity
$N_{0110}$), the second one by states $|0\rangle_a|0\rangle_b$ and
$|2\rangle_a|2\rangle_b$ (negativity $N_{0220}$) and the third one
by states $|1\rangle_a|1\rangle_b$ and $|2\rangle_a|2\rangle_b$
(negativity $N_{1221}$). We note that these subsystems are not
independent -- each of the two-mode states is involved in two
considered \textit{qubit-qubit} subsystems. This also means that
there occur correlations in the presence of entanglement in these
subsystems.

This is illustrated in Fig.~2, where the negativities $N_{0110}$,
$N_{0220}$, and $N_{1221}$ for the subsystems are plotted.
Negativity $N_{0220}$ oscillates with a relatively small amplitude
(around the value of $0.17$) similarly as negativity $N_{1221}$
(having the amplitude around $ 0.3$). On the other hand, values of
negativity $N_{0110}$ show that the system can be treated as a
generator of Bell states in certain time instants in which
$N_{0110} \approx 1$. The generated Bell states are of the form:
\begin{eqnarray}  % 11
 |B_1\rangle&=&\frac{1}{\sqrt{2}}\left(|0\rangle_a|0\rangle_b+i|1\rangle_a|1\rangle_b\right),\nonumber\\
 |B_2\rangle&=&\frac{1}{\sqrt{2}}\left(|0\rangle_a|0\rangle_b-i|1\rangle_a|1\rangle_b\right).
\end{eqnarray}
Moreover, whenever negativity $N_{0110}$ achieves the shallower
minimum (with amplitude approx.$\sim 0.3$), entanglement is
partially transferred to the subsystem spanned by states
$|1\rangle_a|1\rangle_b$ and $|2\rangle_a|2\rangle_b$ $\{1221\}$.
In addition, when negativity $N_{0110}$ equals zero negativity
$N_{1221}$ also vanishes. However, the whole system is not in a
pure state in this specific case because $N_{0220}\neq 0$. The
values of negativity $N_{1221}$ are not high for a fixed value of
$g$, but we have observed that the mechanism of entanglement
transfer alters as $g$ increases. At larger values of $ g $
temporal evolution of negativities $N_{0110}$, $N_{0220}$, and
$N_{1221}$ is more complex (see Fig.~3a). In particular, the role
of higher populated two-mode states ($|2\rangle_a|2\rangle_b$) is
more pronounced and entanglement in subsystem composed of states
$|0\rangle_a|0\rangle_b$ and $|1\rangle_a|1\rangle_b$ is weaker.
There occur time instants in which negativity $N_{0110}$ achieves
one of its minima and nearly simultaneously entanglements in other
both subspaces reach their maximal values. For comparison, we have
drawn in Fig.~3 also parameter $ R $ that quantifies the violation
of Cauchy-Schwartz inequality (CSI, for definition, see later) and
serves as a measure of non-classicality of the whole system.

Inspection of curves in Fig.~3 indicates that maximum of parameter
$ R $ can be reached in time instants in which qubit-qubit
negativities are far from their maximal values. This reflex
complexity of quantum states found in temporal evolution of the
discussed system. It should be noted that, for the
case discussed here, we have assumed the greater value of
parameter $g$. The used value $g=0.6$ has been found optimal for
the entanglement generation within the considered
\textit{qubit-qubit} subsystem. However, the states involving more
than two photons are non-negligibly populated and so we cannot
apply the analytical formulas derived above.

\section{Entanglement generation in damped reservoir}

As any real physical system is not completely isolated, we should
include into considerations about entanglement formation and its
time-evolution also environmental effects. We describe damping in
the standard Born and Markov approximations. We thus apply the
following master equation for the reduced statistical operator $
\hat{\rho} $:
\begin{eqnarray}   % 12
\frac{d}{dt} \hat{\rho} &=&
 -i(\hat{H}\hat{\rho}-\hat{\rho}\hat{H}) + \gamma_a\left[
 \hat{a}\hat{\rho} \hat{a}^\dagger -\frac{1}{2}\left(\hat{\rho} \hat{a}^\dagger
 \hat{a} + \hat{a}^\dagger \hat{a}\hat{\rho} \right)\right] \nonumber \\
 & & \mbox{} +\bar{n}_a\gamma_a\left[\hat{a}^\dagger\hat{\rho}
 \hat{a} + \hat{a}\hat{\rho} \hat{a}^\dagger - \hat{a}^\dagger \hat{a}\hat{\rho} - \hat{\rho} \hat{a}
 \hat{a}^\dagger \right]\nonumber\\
 & & \mbox{} + \gamma_b\left[ \hat{b}\hat{\rho} \hat{b}^\dagger-\frac{1}{2}\left(\hat{\rho} \hat{b}^\dagger \hat{b}
  + \hat{b}^\dagger \hat{b}\hat{\rho}\right)\right] \nonumber \\
 & & \mbox{}
  +\bar{n}_b\gamma_b\left[ \hat{b}^\dagger\hat{\rho} \hat{b}
  + \hat{b}\hat{\rho} \hat{b}^\dagger - \hat{b}^\dagger \hat{b}\hat{\rho} -\hat{\rho}\hat{b}
  \hat{b}^\dagger\right],
\label{master_1}
\end{eqnarray}
where $\gamma_{a,b}$ are damping constants and $\bar{n}_{a,b}$
denote mean photon numbers of noise in modes $a$ and $b$.

Operator equation written in Eq.~(\ref{master_1}) can be
transformed into the following differential equations of motions
for the elements $\rho_{n\,m,k\,l}$ of statistical operator; $
\rho_{n\,m,k\,l}\equiv\langle k|\langle n|\rho|m\rangle|l\rangle
$:
\begin{eqnarray}  % 13
 \frac{d}{dt}\rho_{n\,m,k\,l} &=& -\frac{1}{2}\Bigl[i\chi_a\left(n(n-1)-m(m-1)\right)
  \nonumber \\
 & & \hspace{-1cm} \mbox{} +i\chi_b\left(k(k-1)-l(l-1)\right) \nonumber\\
 & & \hspace{-1cm} \mbox{} + \gamma_a\left(n+m-2\bar{n}_a(n+m+1)\right) \nonumber \\
 & & \hspace{-1cm} \mbox{} + \gamma_b\left(k+l-2\bar{n}_b(k+l+1)\right)\Bigr]\rho_{n\,m,k\,l} \nonumber\\
 & & \hspace{-1cm} \mbox{} + g\sqrt{nk}\;\,\rho_{n-1\,m,k-1\,l} \nonumber \\
 & & \hspace{-1cm} \mbox{} - g\sqrt{(m+1)(l+1)}\;\,\rho_{n\,m+1,k\,l+1} \nonumber\\
 & & \hspace{-1cm} \mbox{} + g^{\star}\sqrt{(n+1)(k+1)}\;\,\rho_{n+1\,m,k+1\,l} \nonumber\\
 & & \hspace{-1cm} \mbox{} - g^{\star}\sqrt{ml}\;\,\rho_{n\,m-1,k\,l-1} \nonumber\\
 & & \hspace{-1cm} \mbox{} + \gamma_a\Bigl(1+\bar{n}_a)\sqrt{(n+1)(m+1)}\;\,\rho_{n+1\,m+1,k\,l}  \nonumber\\
 & & \hspace{-1cm} \mbox{} + \bar{n}_a\sqrt{nm}\;\,\rho_{n-1\, m-1,k\,l}\Bigr] \nonumber\\
 & & \hspace{-1cm} \mbox{} + \gamma_b\Bigl[(1+\bar{n}_b)\sqrt{(k+1)(l+1)}\;\,\rho_{n\,m,k+1\,l+1}  \nonumber\\
 & & \hspace{-1cm} \mbox{} + \bar{n}_b\sqrt{kl}\;\,\rho_{n\,m,k-1\,l-1}\Bigr].
\label{master_el}
\end{eqnarray}
In numerical calculations, the number of equations in
Eq.~(\ref{master_el}) that have to be considered depends strongly
on the value of constant $ g $: the larger the value of $ g $ the
greater the number of equations.

\subsection{Zero-temperature reservoir}

The zero-temperature reservoir is defined by the requirement that
mean number of photons is zero, i.e. the system is influenced only
by vacuum fluctuations of the environment. In the analysis we
restrict ourselves to the subspace that contains states with no
more than two photons in mode $ a $ and $ b $. We have found for
the case discussed here that some amount of entanglement described
by negativities occurs in the system regardless on the damping
strengths $ \gamma_{a,b} $. It is documented in Fig.~4 where the
map of negativity $ N_{1221} $ as a function of time $ t $ and
damping constant $\gamma$ is plotted ($ \gamma \equiv \gamma_a =
\gamma_b $).

A typical temporal evolution of negativity is such that after
oscillations at the beginning an asymptotic nonzero value is
reached (set the inset in Fig.~4). It is interesting to note that
the asymptotic value does not depend on the level of damping (for
$g=0.6$, $ N_{1221} \sim 0.054$). On the other hand, negativity $
N_{0110} $ shows that the effect of \textit{sudden death} of
entanglement occurs in the subsystem with zero and one photons in
modes $ a $ and $ b $. Even more interestingly, negativity $
N_{0220} $ of the subsystem composed of states $ |0\rangle_a $, $
|2\rangle_a $, $ |0\rangle_b $, and $ |2\rangle_b $ brings
evidence of the effects of sudden death and {\textit sudden birth}
of entanglement. After entanglement rebirth negativity $ N_{0220}
$ reaches its nonzero asymptotic value ($ N_{0220} \sim 0.035$ for
$g=0.6$) regardless of the damping strength $ \gamma $. Positions
of the borders between zero and non-zero values of negativities $
N_{0110} $ and $ N_{0220} $ are plotted in Fig.~5. It is worth
mentioning that entanglement described by negativity $ N_{0220} $
lives almost two times longer before its sudden death compared to
that described by negativity $ N_{0110} $.

Analytical treatment enables to reveal the origin of the observed
features to certain extent. If the system is not damped, the
generated states are in a coherent superposition of states $
|i\rangle_a|i\rangle_b $ containing $ i $ photon pairs [compare
the form of Hamiltonian $H_{\rm int}$ in Eq.~(\ref{hamilt})]. In a
damped system, depopulation channels due to spontaneous emission
are opened and, as a consequence, the system has to be described
by statistical operator $ \hat{\rho} $. Nevertheless, projected
statistical operators $ \hat{\rho}^{\rm red} $ of the considered
qubit-qubit subsystems attain an ''X'' type form:
\begin{equation}   % 14
 \hat{\rho}^{\rm red} = \left(
   \begin{array}{cccc}
   a_{11} & 0 & 0 & a_{14} \\
   0 & a_{22} & 0 & 0 \\
   0 & 0 & a_{33} & 0 \\
   a_{41} & 0 & 0 & a_{44}
  \end{array} \right) .
 \label{14}
\end{equation}
In Eq.~(\ref{14}) elements $a_{ii}$ describe populations of the
corresponding states whereas $a_{ij}$ ($i\neq j$) stand for
coherences (probability flows).

The statistical operator $ \hat{\rho}^{\rm red} $ written in
Eq.~(\ref{14}) can be after partial transposition easily
diagonalized. Its eigenvalues can be written as:
\begin{eqnarray}  % 15
 \lambda_1&=&a_{11}, \nonumber\\
 \lambda_2&=&\frac{1}{2}\left[a_{22}+a_{33}-\sqrt{\left(a_{22}-a_{33}\right)^2+4a_{14}a_{41}}\right],
  \nonumber\\
 \lambda_3&=&\frac{1}{2}\left[a_{22}+a_{33}+\sqrt{\left(a_{22}-a_{33}\right)^2+4a_{14}a_{41}}\right],
 \nonumber\\
 \lambda_4&=&a_{44}.
\label{15}
\end{eqnarray}
Nonzero negativity $ N $ occurs whenever at least one of the
eigenvalues written in Eq.~(\ref{15}) is negative \cite{Hill1997}.
The analysis of the formulas in Eq.~(\ref{15}) shows that this
situation is found whenever $a_{22}a_{33}<a_{14}a_{41}$. The
equality $a_{22}a_{33}=a_{14}a_{41}$ then represents the border
condition that localizes the effects of sudden death and sudden
birth of entanglement. A detailed discussion of the relation
between the form of a statistical operator and conditions for
entanglement vanishing or rebuilding can be found in the review
article \cite{F10}.

In detail, negativity $ N_{0110} $ is larger than zero provided
that
\begin{equation}  % 16
 \frac{\rho_{00,11}\rho_{11,00}}{\rho_{01,01}\rho_{10,10}}<1 .
\label{FG1}
\end{equation}
Similarly, the conditon
\begin{equation}  % 17
 \frac{\rho_{00,22}\rho_{22,00}}{\rho_{02,02}\rho_{20,20}}<1
\label{FG2}
\end{equation}
guarantees nonzero values of negativity $ N_{0220} $. In general
and roughly speaking, the conditions written in Eqs.~(\ref{FG1})
and (\ref{FG2}) say that entanglement is lost at the moment when
populations become larger than coherences. However, if coherences
become larger than populations later, entanglement is revealed.
This is documented in Fig.~6 for the evolution of entanglement in
the subsystems described by negativities $ N_{0110} $ and $
N_{0220} $. Whereas the curves in Fig.~6 clearly identify the
instant of sudden death of entanglement in case of negativity $
N_{0110} $ (Fig.~6a), instants of both the sudden death and sudden
birth of entanglement can be determined in case of negativity $
N_{0220} $ (Fig.~6b).

We note that the explanation of occurrence of entanglement sudden
death and sudden birth presented here is very similar to that
discussed by Ficek and Tana\'s in \cite{FT06} when investigating
two two-level atoms in a cavity. They have shown that dark periods
of entanglement are observed provided that the population of a
symmetric state became larger than the appropriate coherences.

We would like to stress that, in our model, entanglement does not
vanish in the long-time limit even for a damped system. Some parts
of the whole system are disentangled as evidenced, e.g., by zero
asymptotic values of negativity $ N_{0110} $. However, there
exist parts of the whole system that remain asymptotically
entangled. Moreover, the study of temporal dynamics of
entanglement has revealed that entanglement can ``flow'' among
different subsystems. This may preserve entanglement from damping
processes and its loss. We also note that the studied system does
not evolve into a pure state which distinguishes it from many
other systems that exhibit total disentanglement in the long-time
limit.

\subsection{Thermal reservoir damping}

Considering a thermal reservoir at finite temperature, mean
numbers $\bar{n}_a$ and $\bar{n}_b$ of reservoir photons are
nonzero and so dynamics of the system is governed by the master
equation in its general form given by Eq.~(\ref{master_1}).
Unfortunately, such reservoirs lead to weakening of entanglement
compared to the case of zero-temperature reservoirs discussed
above.

Assuming smaller values of mean reservoir photon numbers
$\bar{n}_a$ and $\bar{n}_b$, entanglement is generated in all
three considered qubit-qubit subsystems, as described by
negativities $ N_{0110} $, $ N_{0220} $, and $ N_{1221} $ for
finite times. Whereas entanglement indicated by negativities $
N_{0110} $ and $ N_{0220} $ arrives at the crucial points of
sudden death, certain amount of asymptotic entanglement is
preserved as evidenced by nonzero asymptotic values of negativity
$ N_{1221} $. It holds that the larger the mean reservoir
photon-numbers $\bar{n}_a$ and $\bar{n}_b$, the smaller the
asymptotic values of negativity $ N_{1221} $ (see Fig.~7).

The dynamics of entanglement is in general more complex for
finite-temperature reservoirs compared to those at zero
temperature, as documented in Fig.~7 where many periods of
entanglement sudden death and sudden birth occur in the evolution
of negativity $ N_{1221} $. We have also observed, that
entanglement described by negativity $ N_{0110} $ disappears
approximately two times faster compared to that monitored by
negativity $ N_{0220} $. The instants of sudden death of
entanglement depend strongly on damping constants $ \gamma_a $ and
$ \gamma_b $: the larger the values of constants $ \gamma_a $ and
$ \gamma_b $, the sooner the effect of sudden death occurs.
Qualitatively the same dependence as that depicted in Fig.~5 for
zero-temperature reservoirs has been revealed.

Asymptotic values of negativity $ N_{1221} $ are zero for
sufficiently large values of mean reservoir photon numbers
$\bar{n}_a$ and $\bar{n}_b $. In this case, entanglement in the
whole system is completely lost. In this case, also parameter $ R
$ describing violation of Cauchy-Schwartz inequality is smaller
than one indicating classical behaviour of two modes.

We note that, similarly as in the case of zero-temperature
reservoirs, entanglement is present whenever populations are
larger than coherences.

Comparison with the model of a nonlinear coupler interacting with
a thermal reservoir and excited by an external coherent field as
analysed in \cite{KL09} underlines a distinguished property of
this system - the ability to generate asymptotic entanglement at
finite temperatures.

\section{Nonclassical correlations of integrated intensities}

Entanglement between fields in modes $ a $ and $ b $ has been
monitored using negativity of different qubit-qubit subsystems
defined inside modes $ a $ and $ b $. The presence of entanglement
in at least one of the subsystems reflected entanglement between
optical fields in these modes. We have observed that whenever the
system shows entanglement also nonclassical correlations between
integrated intensities of fields in modes $ a $ and $ b $ occur
\cite{Perina2009}. These nonclassical correlations of intensities
occur only provided that the joint quasi-distribution (in the form
of a generalized function) of integrated intensities related to
normal ordering of field operators attains negative values for
some regions of intensities \cite{P91}. These negative values can
be monitored, e.g., using second-order intensity correlation
functions $ \Gamma^{(2)} $ that violate the Cauchy-Schwartz
inequality. Violation of this inequality can be quantified using
parameter $ R $ defined a \cite{LYXLY06}:
\begin{equation}   % 18
 R(t)=\frac{\Gamma^{(2)}_{a,b}(t)}{\sqrt{\Gamma^{(2)}_{a,a}(t)\Gamma^{(2)}_{b,b}(t)}},
\label{R}
\end{equation}
where the second-order intensity correlation functions $
\Gamma^{(2)} $ are defined as
\begin{equation}  % 19
 \Gamma^{(2)}_{k,l}(t)= {\rm Tr} \left\{ \hat{\rho}(t) {\cal N} \hat{I}_k
  \hat{I}_l\right\} .
\label{19}
\end{equation}
In Eq.~(\ref{19}), $ \hat{\rho} $ means statistical operator, $
\hat{I}_k = \hat{k}^\dagger \hat{k} $ and $ {\cal N} $ stands for
normal-ordering operator. Nonclassical states obey the inequality
$ R > 1 $.

Considering the system without damping, fields in modes $ a $ and
$ b $ are always entangled and also parameter $ R $ is always
larger than one (compare temporal evolution of negativities $ N $
and parameter $ R $ in Fig.~3a). Considering zero-temperature
reservoirs, nonclassical correlations of intensities are always
observed. On the other hand, finite-temperature reservoirs
gradually destroy nonclassical correlations of intensities,
similarly as they weaken entanglement. Nonclassical correlations
in intensities are asymptotically preserved for smaller values of
mean reservoir photon numbers $ \bar{n}_a $ and $ \bar{n}_b $.
However, larger values of mean photon numbers $ \bar{n}_a $ and $
\bar{n}_b $ result in the loss of nonclassical correlations in
finite times. We also note that the greater the values of damping
constants $ \gamma_a $ and $ \gamma_b $ the sooner the
nonclassical correlations are lost. These features are documented
in Fig.~8, where temporal evolution of parameter $ R $ is plotted.

Investigation of entanglement based on negativities of qubit-qubit
subsystems composed of Fock states with low numbers need not
guarantee a correct determination of non-classicality of intensity
correlations. The problem is that entanglement ``flows'' among
different subsystems during its temporal evolution and it may
happen that it exists only in subspaces composed of Fock states
with larger numbers. As an example we consider systems described
in Fig.~8 ($\bar{n}_a=\bar{n}_b=0.4$, $g=0.6$) and exhibiting
entanglement also in subsystems involving states
$|3\rangle_a|3\rangle_b$ and $|4\rangle_a|4\rangle_b$.

\section{Conclusions}

We have analysed entanglement between number states generated in
two nonlinear Kerr oscillators pumped by optical parametric
process using qubit-qubit negativities. We have shown that
maximally-entangled states of Bell type can be generated under
certain conditions. Interaction of the system with
zero-temperature reservoirs on one side weakens the ability to
generate entanglement, but on the other side leads to a more
complex evolution of entanglement with the effects of its sudden
death and sudden birth. The effect of sudden birth occurs provided
that coherences in the system dominate populations. Instants of
sudden deaths and sudden births of entanglement differ for
different qubit-qubit systems which reflects ``dynamical flow of
entanglement'' in the system. Finite reservoir temperatures
inhibit the effect of entanglement sudden birth and destroy
asymptotically entanglement for greater reservoir mean photon
numbers. We have found that entanglement occurs whenever there
exist nonclassical correlations in intensities of two oscillator
modes that violate Cauchy-Schwartz inequality.

\acknowledgments J.P. thanks the support from projects
IAA100100713 of GA AS CR and COST OC 09026 of the Ministry of
Education of the Czech Republic.

\newpage
\begin{figure}  % fig. 1
 \begin{center}
 \vspace*{-0.1cm}
 \resizebox{0.8\hsize}{!}{\includegraphics{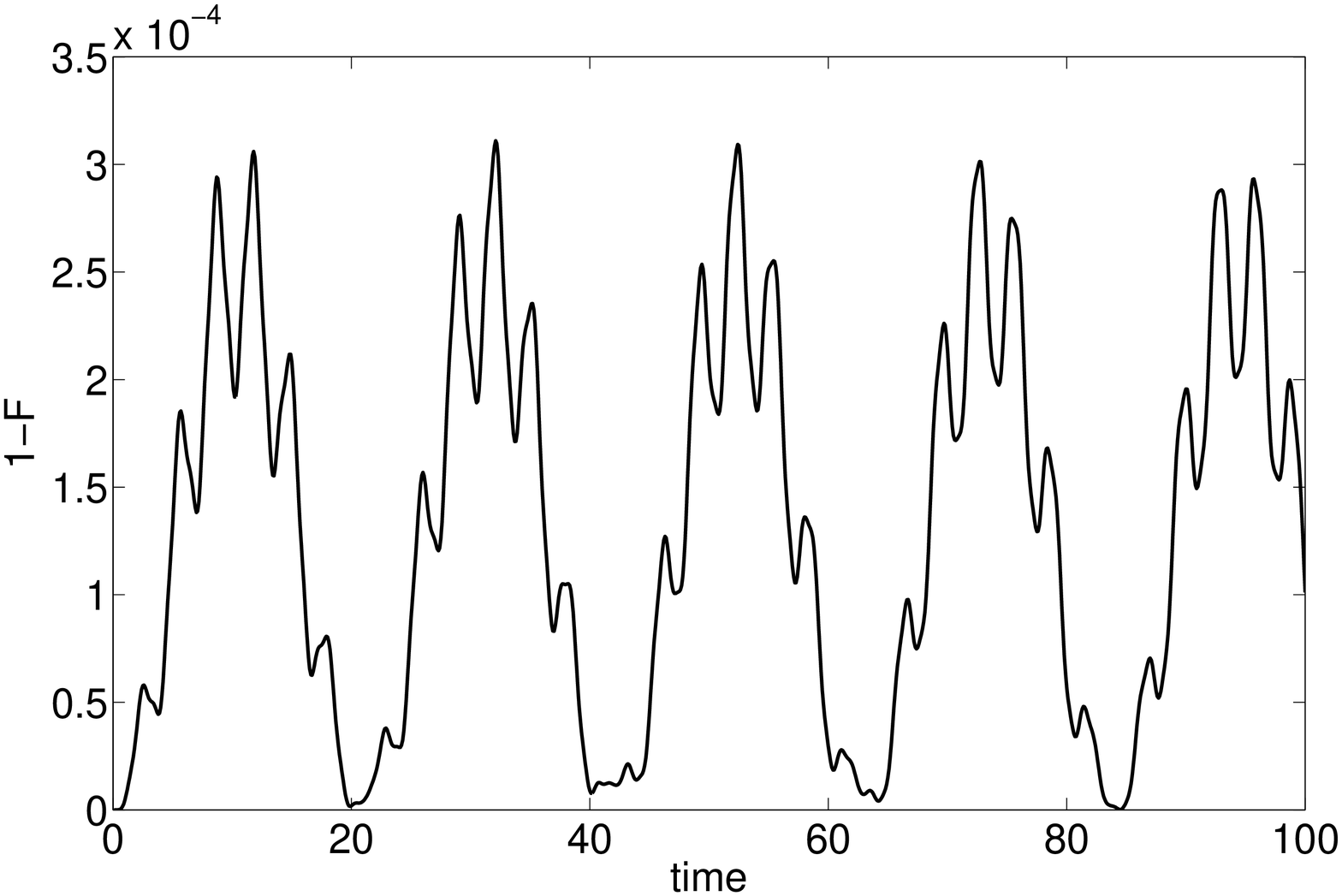}}
 \vspace*{-0.5cm}
 \caption{Declination of fidelity $ F $ from unity as a function of scaled time $ t $.
 Fidelity $ F $ is calculated between the state defined in the subspace spanned by
 states $|0\rangle_a|0\rangle_b $, $ |1\rangle_a|1\rangle_b $,
 and $ |2\rangle_a|2\rangle_b$ and the state evolving in the subspace
 containing states $|i\rangle_a|j\rangle_b$ for $i,j=0, \ldots,10 $. The scaled time $ t $
 is measured in $1/\chi_{a,b}$ units. Initial vacuum state $|0\rangle_a|0\rangle_b$ is assumed,
 $g=0.15$.}
\end{center}
\end{figure}

\begin{figure}  % fig. 2
 \begin{center}
  \vspace*{-0.1cm}
  \resizebox{0.8\hsize}{!}{\includegraphics{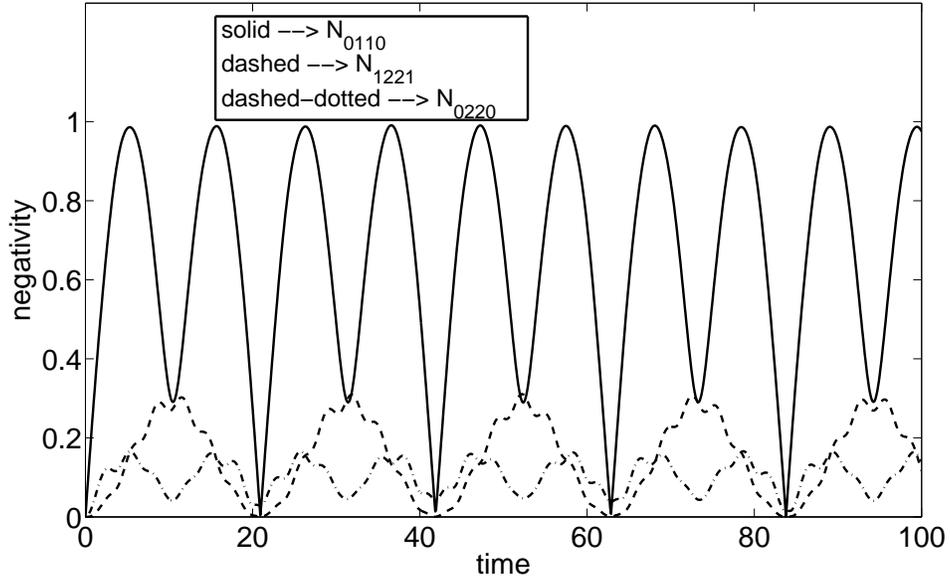}}
  \vspace*{-0.5cm}
  \caption{Negativities $N_{0110}$, $N_{0220}$, and $N_{1221}$ as functions of scaled
  time $ t $ measured in $1/\chi_{a,b}$ units. The coupling $g=0.15$, other parameters and used units are the same as in caption to Fig.~1.}
\end{center}
\end{figure}

\begin{figure}   % fig. 3
 \begin{center}
  \vspace*{-0.1cm}
  \resizebox{0.8\hsize}{!}{\includegraphics{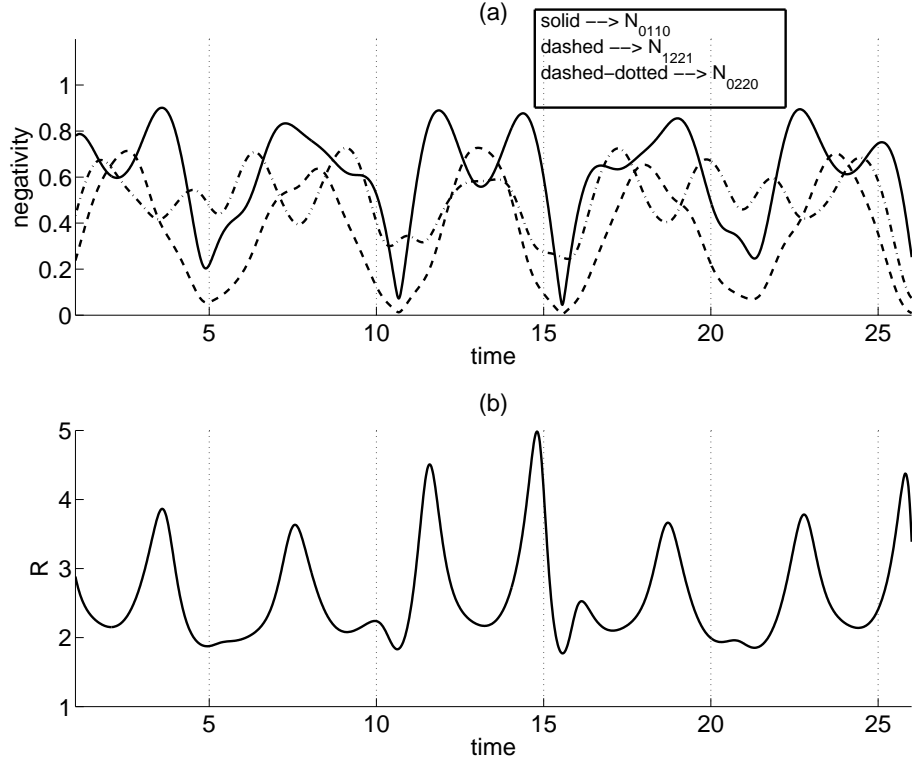}}
  \vspace*{-0.5cm}
  \caption{(a) Negativities $N_{0110}$, $N_{0220}$, and $N_{1221}$ and
  (b) parameter $R$ indicating violation of the
  Cauchy-Schwartz inequality violation as they depend of
  scaled time $ t $. The scaled time $ t $
 is measured in $1/\chi_{a,b}$ units. Initial vacuum state $|0\rangle_a|0\rangle_b$ is assumed, $g=0.6$.}
 \end{center}
\end{figure}

\begin{figure}   % 4
 \begin{center}
  \vspace*{-0.1cm}
  \resizebox{0.8\hsize}{!}{\includegraphics{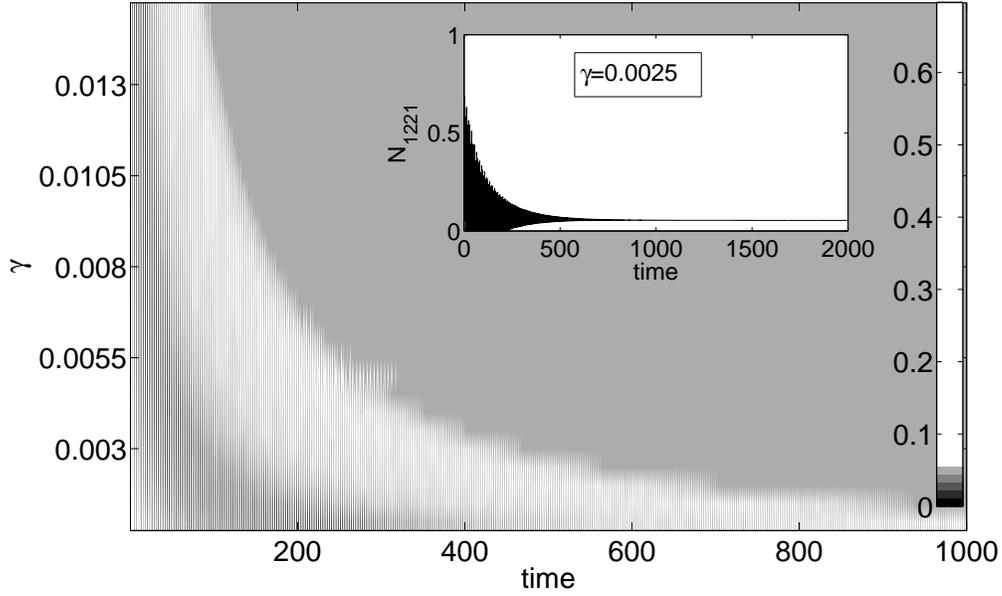}}
  \vspace*{-0.5cm}
  \caption{Map of negativity $N_{1221}$ as a function of scaled in $1/\chi_{a,b}$ units time $ t $
  and damping $\gamma $ (measured in $\chi_{a,b}$ units). Nonzero values of negativity $N_{1221}$ occur inside
  grey and black areas. In the inset temporal evolution of negativity $N_{1221}$
  for $ \gamma = 0.0025 $ is shown. The coupling $g=0.6$, other parameters and units are the same as in caption to Fig.~3.}
\end{center}
\end{figure}

\begin{figure}   % fig. 5
 \begin{center}
  \vspace*{-0.1cm}
  \resizebox{0.9\hsize}{!}{\includegraphics{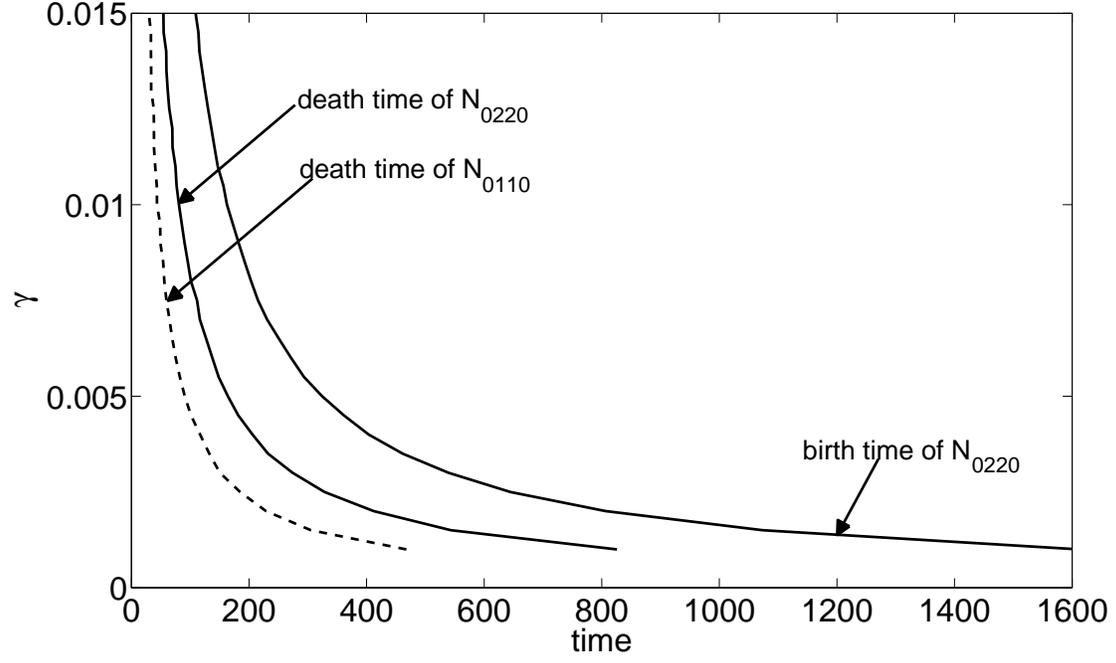}}
  \vspace*{-0.5cm}
  \caption{Borders between the regions with zero/nonzero
  negativities $N_{0110} $ (dashed curve) and $ N_{0220} $ (solid curves)
  at the time axis as they depend on damping $ \gamma $  (measured in $\chi_{a,b}$ units). The curves indicate
  instants in which the effects of sudden death and sudden birth of entanglement occur.
  In the area between two solid lines, no entanglement described
  by negativity $ N_{0220} $ is found.
  The coupling $g=0.6$, other parameters and the units used here are the same as for Fig.~3.}
\end{center}
\end{figure}

\begin{figure}  % figs. 6a and 6b
 \resizebox{0.8\hsize}{!}{\includegraphics{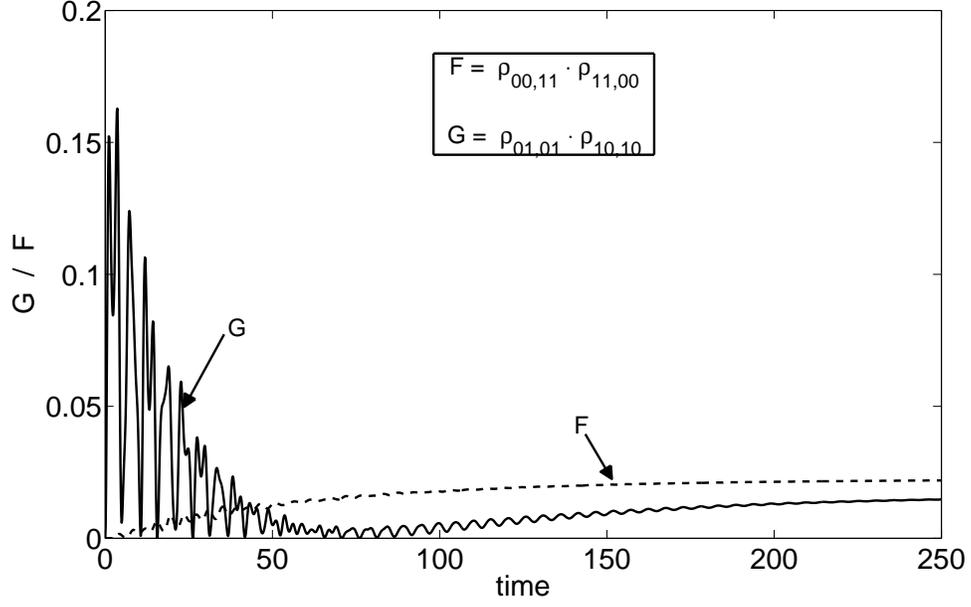}}
 \resizebox{0.8\hsize}{!}{\includegraphics{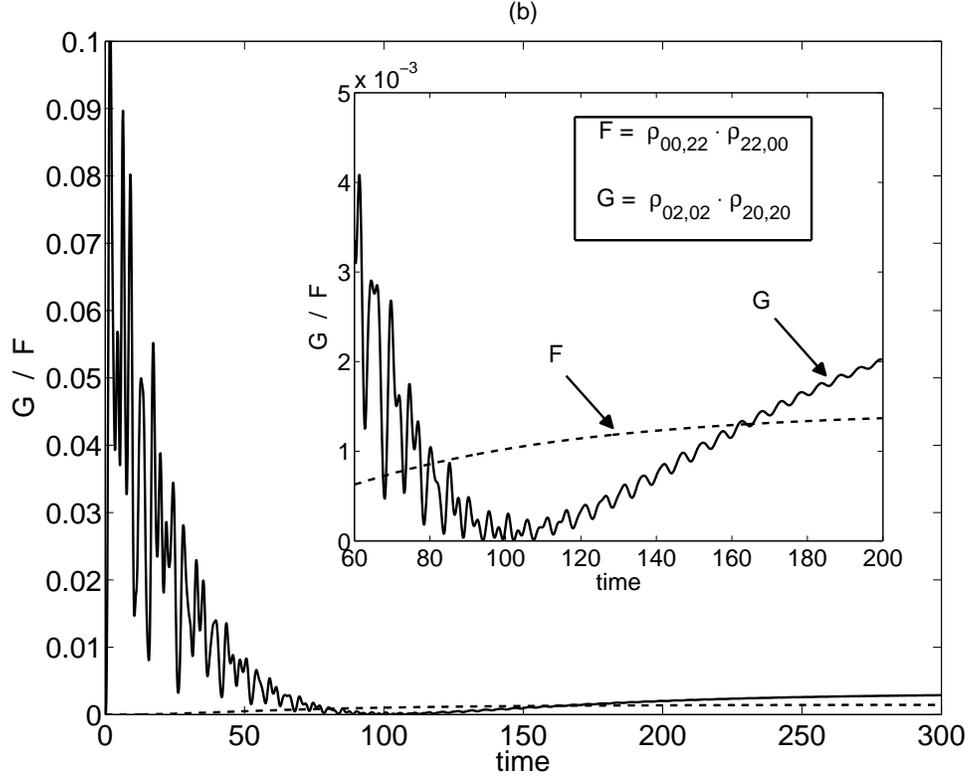}}
 \caption{Time evolution of quantities $ F $ ($ F = \rho_{00,11}\rho_{11,00}
 $) and $ G $ ($ G = \rho_{01,01}\rho_{10,10} $) giving numerators
 and denominators, respectively, in conditions written in
 Eqs.~(\ref{FG1}) (a) and Eqs.~(\ref{FG2}). Crossing of the curves
 indicate instants of sudden death and sudden birth of
 entanglement. Inset in (b) shows a detail of temporal evolution;
 $\gamma=0.01$. The coupling $g=0.6$, other other parameters are the same as in caption
 to Fig.~3. Time $ t $ is measured in $1/\chi_{a,b}$ units.}
\end{figure}

\begin{figure}  % fig. 7
 \begin{center}
  \vspace*{-0.1cm}
  \resizebox{0.9\hsize}{!}{\includegraphics{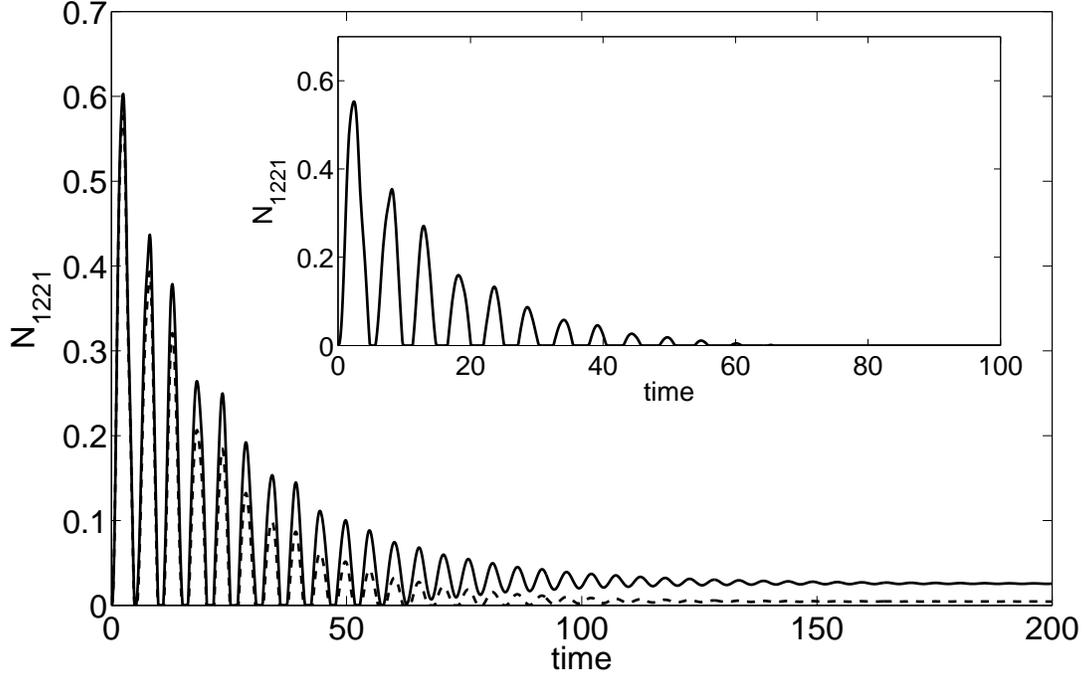}}
  \vspace*{-0.5cm}
  \caption{Time evolution of negativity $N_{1221}$ for
  $ = \equiv n_a=n_b=0.1$ (solid curve), $n=0.2$ (dashed curve),
  and $n=0.3$ (inset); $\gamma=0.01$. The coupling $g=0.6$, other parameters and units are
  the same as in caption to Fig.~3.}
\end{center}
\end{figure}

\begin{figure}   % Fig.8
 \begin{center}
  \vspace*{-0.1cm}
  \resizebox{0.9\hsize}{!}{\includegraphics{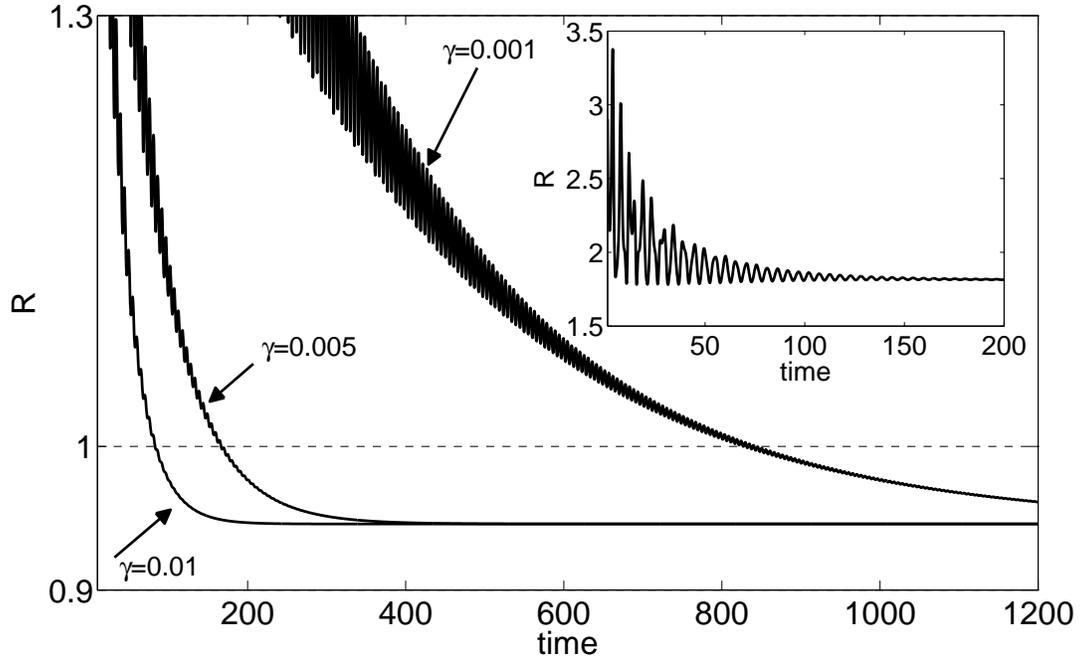}}
  \vspace*{-0.5cm}
  \caption{Temporal evolution of parameter R for damping constants
  $\gamma = 0.001 $, $ 0.005 $, and $ 0.01 $ assuming $\bar{n}=0.4$. For comparison,
  temporal evolution of parameter $ R $ for zero-temperature
  reservoirs ($n=0$) is given for $\gamma=0.01$. The coupling $g=0.6$, Values of other
  parameters are the same as in caption to Fig.~3. Time $ t $ is measured in $1/\chi_{a,b}$ units.}
 \end{center}
\end{figure}

\end{document}